# Hoop Diagrams: A Set Visualization Method

Peter Rodgers[1][0000-0002-4100-3596], Peter Chapman[2][0000-0002-5524-5780], Andrew Blake[3][0000-0001-5856-4544], Martin Nöllenburg[4][0000-0003-0454-3937], Markus Wallinger[4][0000-0002-2191-4413], and Alexander Dobler[4][0000-0002-0712-9726]

[1] University of Kent, UK
[2] Edinburgh Napier University, UK
[3] University of Brighton, UK
[4] Technische Universität Wien, Austria

p.j.rodgers@kent.ac.uk

**Abstract.** We introduce Hoop Diagrams, a new visualization technique for set data. Hoop Diagrams are a circular visualization with hoops representing sets and sectors representing set intersections. We present an interactive tool for drawing Hoop Diagrams and describe a user study comparing them with Linear Diagrams. The results show only small differences, with users answering questions more quickly with Linear Diagrams, but answering some questions more accurately with Hoop Diagrams. Interaction data indicates that those using set order and intersection highlighting were more successful at answering questions, but those who used other interactions had a slower response. The similarity in usability suggests that the diagram type should be chosen based on the presentation method. Linear Diagrams increase in the horizontal direction with the number of intersections, leading to difficulties fitting on a screen. Hoop Diagrams always have a square aspect ratio.

## 1 Introduction

This paper introduces and evaluates a new set visualization technique, Hoop Diagrams. Hoop Diagrams are a circular visualization, where each hoop represents a set, and each sector represents a set intersection. The hoops are broken concentric circular lines, so that the appearance of lines in a sector indicates that the corresponding intersection of sets is non-empty.

To aid our investigation into Hoop Diagrams as a visualization tool, we have designed and implemented an interactive tool for drawing Hoop Diagrams. This allows set and intersection highlighting, reordering of sets and intersections in the diagram, and diagram rotation.

Whilst such set data can also be visualized with Venn and Euler diagrams, the closest current visualization method to Hoop Diagrams are Linear Diagrams. Linear Diagrams have been shown to be an effective visualization method [13][20] hence we concentrate on an empirical and practical comparison between these two diagram types.

Figure 1 illustrates a Hoop Diagram visualising the relationships between six sets. The legend provides the set names in alphabetical order. Each coloured hoop (which follow a circular path) represents one of the six sets. Colour classifies each curve and corresponds to a name on the legend. A black circle on the outside and a smaller circle on the inside form the limits of the hoops. Concentric grey circles delineate the path of each coloured curve. Circle sectors are divided by spokes emanating from the circle centre. Each sector represents a set intersection, so the presence of hoops in a sector means that there is an non-empty intersection of the corresponding sets in the data.

In the examples in this paper, our sets are interests that people may have and the intersections are combination of interests that at least one person has. So that in Figure 1, the first sector reading clockwise (at the 1 O'clock position) has the red and blue lines present, indicating that there is someone that has both the interests "Dogs" and "Poker", but no other interests. The sector immediately to the left (11 O'clock position) has only the brown line present, indicating that there is a person who has only the interest "Esport". There is no sector where the red and purple lines are both present, hence there is no person with only the two interests "Dogs" and "Hifi".

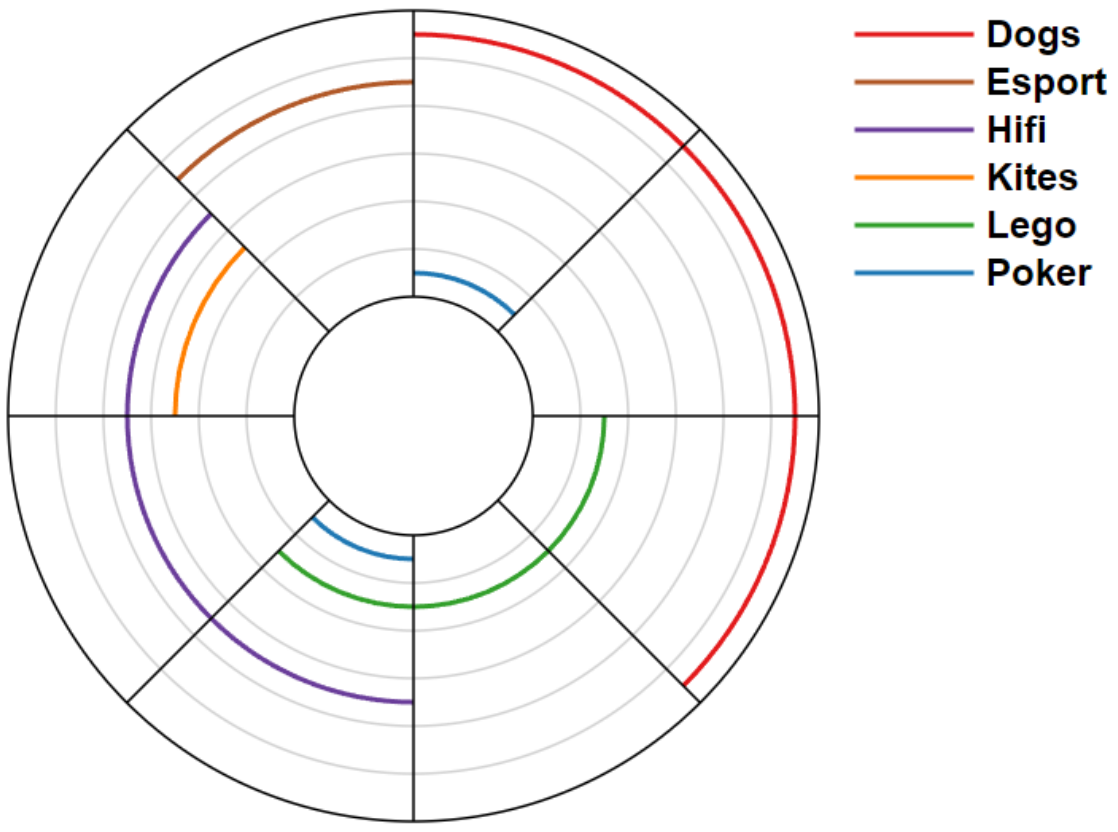

**Fig. 1.** Hoop Diagram.

Figure 2 shows a Linear Diagram visualising the same data as the Hoop Diagram in Figure 1. The set names are shown in alphabetical order on the left. Each row contains a coloured line and represents one of the six sets. Horizontal grey lines delineate the path of each coloured line. Columns are divided by vertical lines. Each column represents a set intersection, so the presence of coloured lines in a column means that there is an non-empty intersection of the corresponding sets in the data.

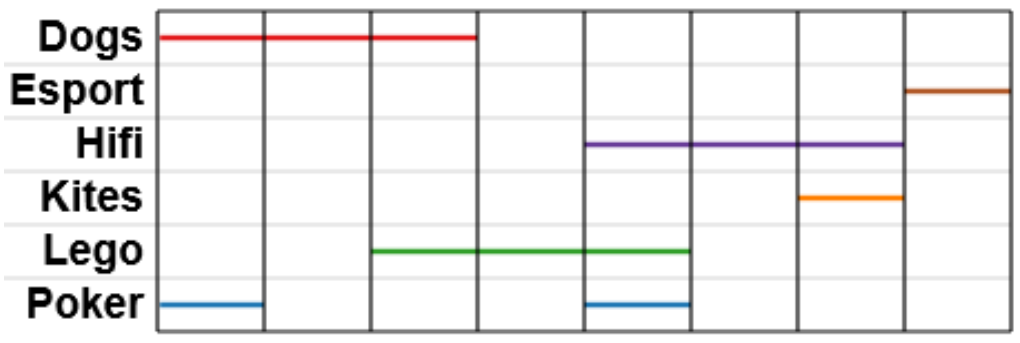

**Fig. 2.** Linear Diagram showing the same data as Figure 1.

We compared both static and interactive versions of Hoop Diagrams against Linear Diagrams. This empirical study shows only small differences in user performance between Hoop and Linear Diagrams. Users were presented with a mix of two types of question: set oriented (requiring the close examination of entire sets) and intersection oriented (requiring the close examination of entire intersections). They answered questions with Linear Diagrams more quickly, but users answered intersection oriented questions with Hoop Diagrams more accurately. We present an exploration of interactions which indicate that those using set order and intersection highlighting were more accurate, but those who used other interactions had a slower response. Our study details, as well as access to interactive Hoop Diagram and Linear Diagram software tools can be found at **https://www.eulerdiagrams.com/hoop/**.

We also explore the usability of these diagrams from a screen real estate perspective. Hoop Diagrams have hoops of different sizes – inner sets use hoops with a smaller radius than outer sets, thus, they use less space and have lower visual prominence in the diagram, which may be seen as a disadvantage compared to Linear Diagrams, which have the same width for all set intersections. Hoop diagrams do not change size as the number of intersections increase, instead reducing the space available for each intersection. As the number of sets increases, Hoop Diagrams grow in both horizontal and vertical directions. This means that the diagrams maintain a square aspect ratio. In contrast, Linear Diagrams grow only in the vertical direction as the number of sets increase and grow in the horizontal direction when the number of intersections increase.

The rest of the paper is organized as follows. Section 2 gives a summary of prior efforts to visualize set based data and gives a theoretical explanation of why Hoop Diagrams might be an effective visualization from a cognitive perspective. Section 3 gives a detailed definition of Hoop Diagrams and explains the interactions available in the software tool. Section 4 outlines our empirical approach. Section 5 gives the results of our comparison between Hoop Diagrams and Linear Diagrams. Section 6 explores the data from the log of interactions. Finally, Section 7 gives our conclusions and discusses future possible work.

## 2 Background

### 2.1 Linear Diagrams

Linear Diagrams can be traced back to Leibniz [8]. Each set is represented as a line drawn in the horizontal direction. The presence of vertically overlapping lines in a column means that the corresponding intersection of sets is not empty. A line may be broken and sets are often represented by multiple line sectors. Set intersections not appearing in the diagram are considered to be empty. See Figure 2 for a example of a Linear Diagram. The method shares similarities to parallel bargrams [24], double decker plots [15] and UpSet [16].

A design study into Linear Diagrams [20] provided evidence that using guidelines, a minimal number of line sectors (a lower number of line breaks) and thin lines for sets led to significantly improved user performance. Given the similarity in the design of Linear Diagrams to Hoop Diagrams, this study informs the design of Hoop Diagrams used in this paper, which follow these guidelines. A study on interactivity in Linear Diagrams [4], found that interactivity improved participants' accuracy, confidence and speed, a result that motivates our use of interactive Hoop and Linear Diagrams.

Theoretical work exists that shows that satisfying one of the design guidelines of linear diagrams – minimizing the number of line segments – is NP-hard in general [6][9]. However, there exist algorithms that achieve the minimum number of line segments for real-world data [9]. Furthermore, a recent paper explores reducing the vertical space required by Linear Diagrams by visualizing multiple sets in the same row of the linear diagram [23]

### 2.2 Other Set Visualization Methods

Venn [22] and Euler [12] diagrams are widely used alternatives to Linear Diagrams. Here, sets are represented by curves enclosing regions. Region overlaps show which set intersections are not empty. Venn and Euler differ in the representation of empty sets. In Venn diagrams, all possible intersections are shown and empty intersections are shaded. In Euler diagrams, empty intersections are not shown. The consequence for Venn diagrams is that the number of intersections is exponential to the number of sets regardless of the number of set intersections present, hence this method has serious scalability issues. The consequence for Euler diagrams of not including empty intersections is that poor wellformedness can often appear. This includes features such as concurrency, triple points and non-simple curves. There is evidence to suggest that breaking such conditions adversely impacts understanding [21]. Limiting the shape of Euler diagrams has been attempted in systems such as RectEuler [19] which represent Euler-like diagrams as rectangles, although studies indicate that simpler shapes such as circles may be more effective for understanding [3]. We note that prior studies indicate Linear Diagrams are more understandable than Euler diagrams [5][13], hence in this paper, we concentrate our comparison between Hoop and Linear Diagrams.

We also note the use of a variety of other set visualization methods, for a survey, see [2]. Those that use lines or region encoding include LineSets [1] and Bubble Sets [7]. However, prior embedding of items contained in the sets is required for these methods as they place lines or regions over existing data items. Node-link techniques, such as PivotPaths [10] also rely on the existence of items, using links between items to indicate shared set membership. Also related are circular visualization methods, see a survey on radial visualizations [11].

# 3 Hoop Diagrams

## 3.1 Hoop Diagram Overview

Hoop Diagrams have sets represented by broken concentric circles ("Hoops"), see Figure 1. Comparing Hoop Diagrams to Linear Diagrams, the same labels and lines are present, as are the intersections, but intersections are represented as columns in Linear Diagrams, rather than sectors as in Hoop Diagrams. Figure 2 shows an equivalent Linear Diagram to the Hoop Diagram in Figure 1. Hoop Diagrams can be seen as transforming into Linear Diagrams by separating the circle at 12 O'clock, straightening the set lines and moving the labels to the left of the diagram.

Compared to Linear Diagrams, we consider the benefit of Hoop Diagrams to be:

1. Compact representation. We consider that the consistent use of two dimensions for intersections provides a more desirable aspect ratio and effective use of display real estate than Linear Diagrams, which suffer when the number of intersections is large, as the diagram becomes unreadable as the intersections get very far from the labels in the X direction. Linear Diagrams scale in the Y direction as the number of sets increase. Hoop diagrams do not increase in size as the number of intersections increases, instead the intersections are presented in a smaller sector of the circle. Hoop diagrams scale linearly in both X and Y directions with the number of sets. This means that the labels, although always separated from the hoops they label, do not get more distant from them.
2. No excessive spacing between intersections. The circular nature of the Hoop Diagram leaves all intersections in close proximity. This contrasts with Linear Diagrams, where the leftmost and rightmost intersections may be a considerable distance apart.

Of course, Linear Diagrams may be considered to have advantages over Hoop Diagrams. We regard the visual simplicity of Linear Diagrams to be the representation's major advantage. In particular, the broken straight line of sets and the column structure of intersections are simpler than the broken circles of sets and sector-based structure of intersections in Hoop Diagrams. Linear Diagrams provide the same space for each set in an intersection, whereas the concentric circles of Hoop Diagrams means there is different spacing for each set, with the lines for inner sets having considerably less circumference than outer sets. Whilst Linear Diagrams can disappear off the screen as intersection numbers increase, for Hoop Diagrams, the size of diagram does not change, but the width of intersections decreases, making discerning the sets present in intersections more difficult with a large number of sets.

## 3.2 Interactions

Here we describe the interactions provided by our web based system. We regard these as a basic set of functionality. More sophisticated interactions, including those that are task dependent, could also be provided by expanding the system.

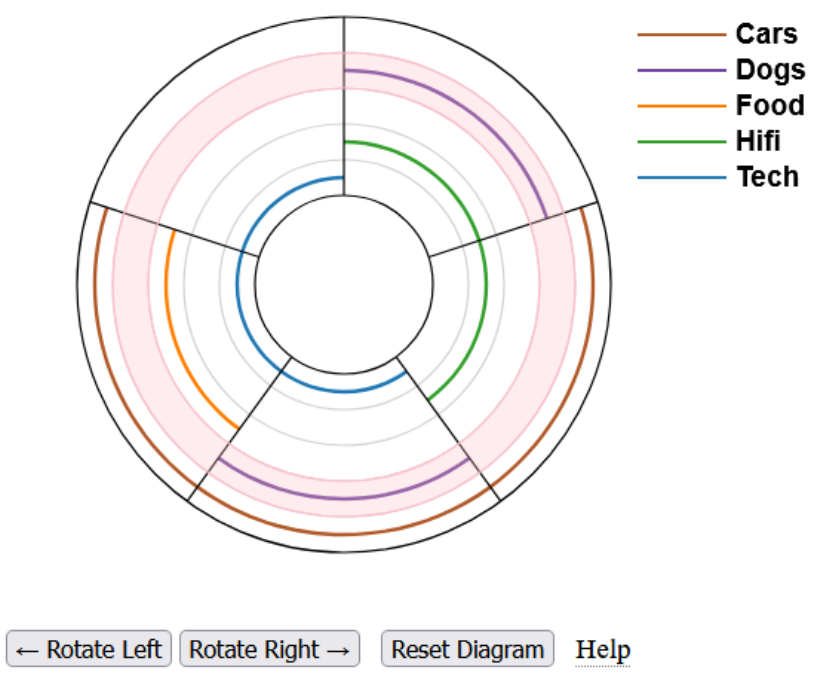

Set highlighting. Set “Dogs” is highlighted.

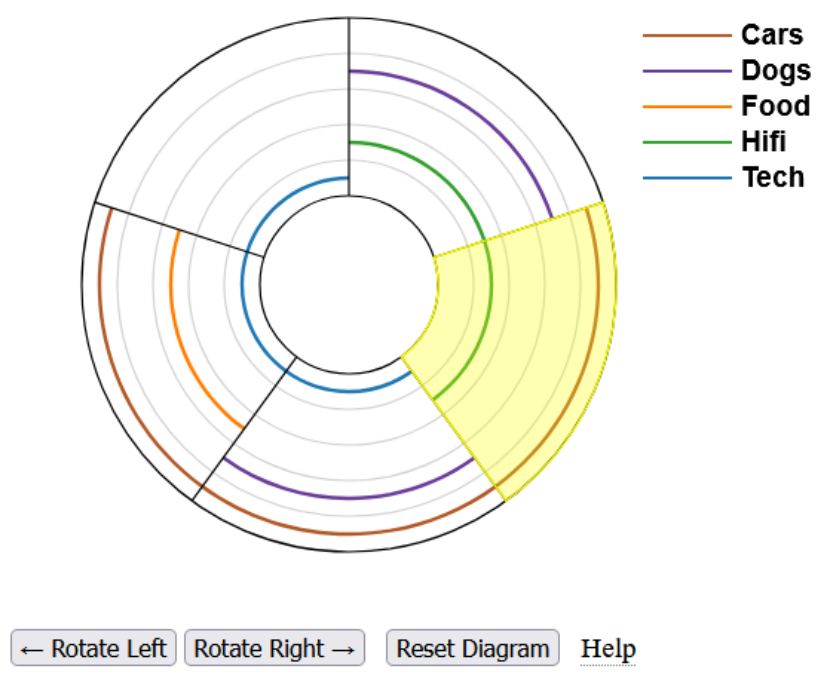

Intersection highlighting. The intersection containing “Cars” and “Hifi” is highlighted.

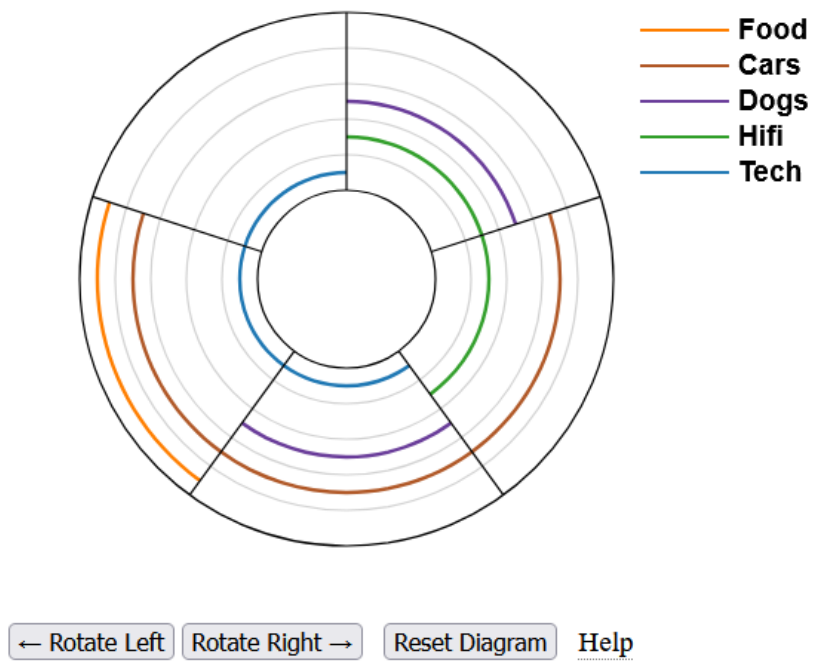

Bring set to outside. “Food” has moved from the third hoop to the outer hoop.

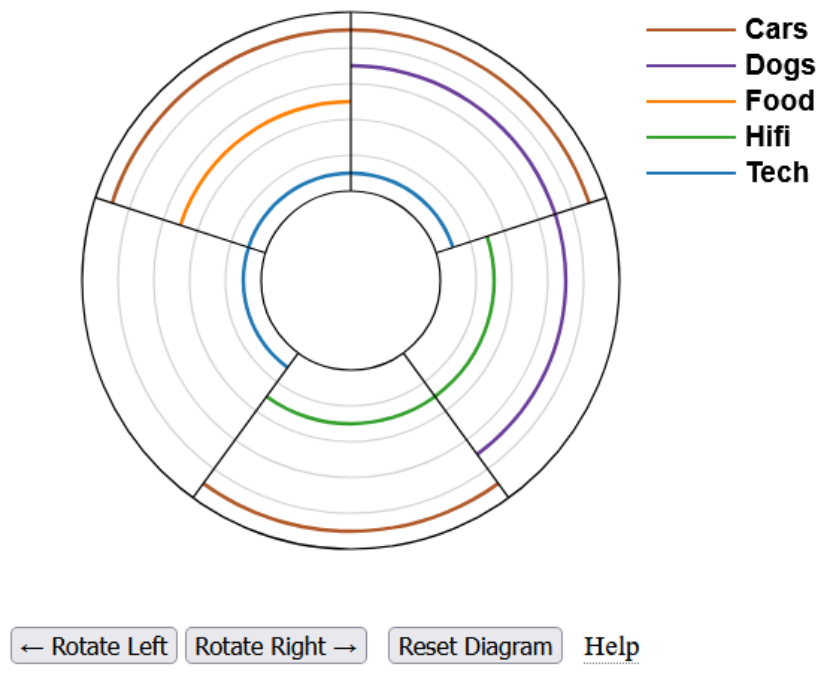

Set reordering. The intersections for set “Dogs” have been brought together.

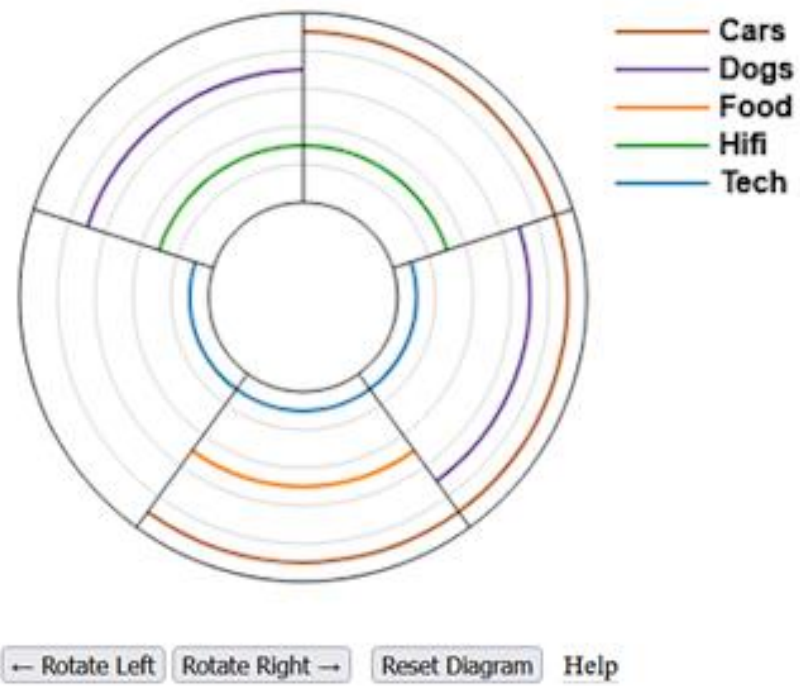

Rotate left. The diagram has been rotated by one sector left compared to the top right diagram.

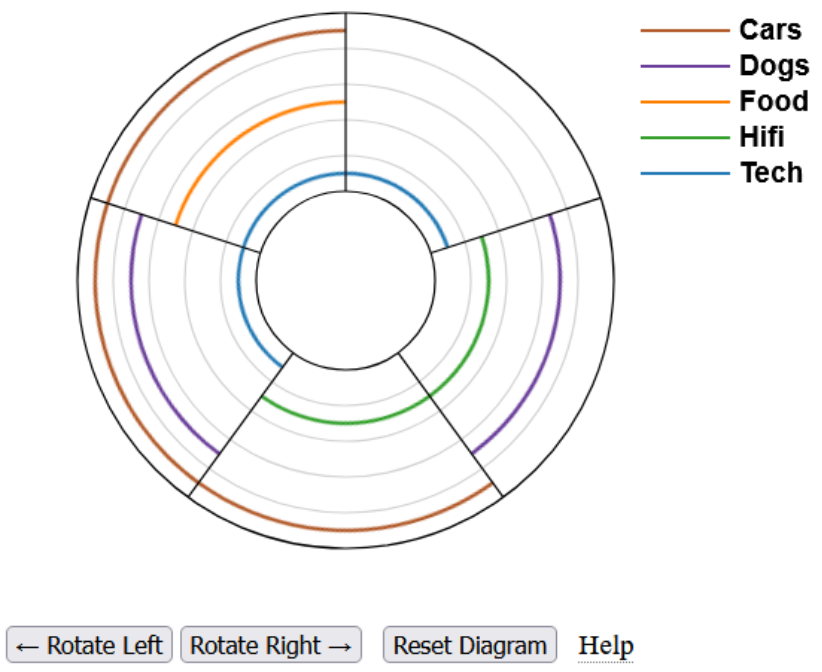

Rotate right. The diagram has been rotated by one sector right compared to the top right diagram.

**Fig. 3.** Layout changes after interaction is applied to the second training diagram.

1. Set Highlighting: Figure 3 top left shows the diagram with a set highlighted by hovering over the hoop. The mouse is in the second hoop from the outside, “Dogs”.

2. Intersection Highlighting. Figure 3 top right shows the diagram with an intersection highlighted by hovering inside or outside the intersection sector. The mouse is just outside the second hoop clockwise, the intersection with “Cars” and “Hifi” only.
3. Bring Set to Outside. Figure 3 middle left shows the diagram after the label “Food” has been clicked by the left mouse button. “Food” is brought to the outside with “Cars” and “Dogs” moved inwards by one hoop.
4. Reorder Set. Figure 3 middle right shows the diagram after the second hoop from the outside, “Dogs” has been clicked by the left mouse button. The sectors are reordered so that all the sectors containing “Dogs” are together, starting from 12 O’clock, meaning the “Dogs” line is continuous. The system attempts to minimize the number of line segments for the other hoops.
5. Rotate. Figure 3 bottom left shows the diagram after the “Rotate Left” button has been pressed once. Figure 3 bottom right shows the diagram after the “Rotate Right” button has been pressed twice after the Rotate Left button was pressed.

The “Reset Diagram” button returns the diagram to its original configuration. The effect on the screen of reordering is animated, using HTML Transitions, with each animation taking 1 second. The reordering interactions are composable, for example, it is possible to have a different set on the outside, along with the set line of choice continuous whilst rotating as many times as desired.

We also developed an interactive tool for Linear Diagrams, to support the studies given in Section 4 where each of the above Hoop Diagram interactions has an equivalent in the Linear Diagram tool:

1. Set Highlighting: A set can be highlighted by hovering over a line.
2. Intersection Highlighting. An intersection can be highlighted by hovering above or below a column.
3. Bring Set to Top. If a label on the left is clicked by the left mouse button, it is raised to the top, with the other labels that were above it lowered by one set.
4. Reorder Set. Clicking on a line reorders the columns so that all the columns for the clicked set are together starting from the left, so that the chosen set line is continuous.
5. Rotate. Pressing a Rotate button rotates the diagram by one column in the desired direction, so “Rotate Left” moves all columns left by one, except the leftmost column that becomes the rightmost column. The “Rotate Right” button moves all columns right by one and the rightmost column becomes the leftmost column.

As with interaction in Hoop Diagrams, the “Reset Diagram” button returns the diagram to its original configuration. All changes are animated.

## 4 Methodology

We defined the following research questions to guide our exploration:
- RQ1: Is data visualized with Hoop Diagrams more understandable than data visualized with Linear Diagrams?
- RQ2: Does interaction in Hoop Diagrams or Linear Diagrams aid understanding?

This led to the following Research Hypotheses related to RQ1:

- H1: Hoop Diagrams are more understandable than Linear Diagrams.
- H2a: Dynamic Linear Diagrams are more understandable than Static Linear Diagrams.
- H2b: Dynamic Hoop Diagrams are more understandable than Static Hoop Diagrams.

RQ2 was to be addressed by a more exploratory approach, where we logged interaction activity to look at:

- Which are the most/least used interactions in Dynamic Hoop/Linear Diagrams?
- Which interactions in Dynamic Hoop/Linear Diagrams are used in successful/unsuccessful tasks?

### 4.1 Data Capture

The study was aimed at both research questions and aimed to answer H1, H2a and H2b, as well as provide data for the exploration of interaction activity.

Participants were recruited through the Prolific platform and redirected to a web based study vehicle hosted at the University of Kent. On completion they were returned to Prolific. All those completing and correctly answering at least one attention check question were paid, although answering both attention check questions correctly was required for the data to be included in our analysis. We calculated the payment depending on expected time for completion (mostly 20 minutes, although some early studies were 25 minutes) and the reward was set at the UK National Living Wage, which at 1 April 2023 was £10.42 an hour, although some studies took place before this for a reward at the UK National Living Wage of £9.50 an hour. The maximum time for completion is assigned by Prolific, and for a 20 minute expected completion the maximum time allowed was 67 minutes. Throughout the studies the median completion time was under the expected time. Participants were distributed through the Prolific standard sample method. We placed prescreening conditions as follows: they must use laptop or desktop computers; participant approval rate must be 99/100; they had a minimum number of 5 prior submissions; and we excluded participants from previous studies on Hoop Diagrams. Participants passed from prolific would be assigned to each condition in turn in an attempt to keep condition completion relatively even. This is a random assignment, as we made no use of participant information when making condition assignments. The completion per condition ratio was not always even as some participants would return the study before completion or would time out. We ran studies in varying size batches of participants to control load on the web servers and make an attempt to even up condition completion ratios towards the end of the study by removing overrepresented conditions from the rotation.

Our diagrams were formed from real-world Twitter Circles data, obtained from the SNAP data set collection [18]. We used the same input data to form the diagrams for each condition of each study. For each Twitter Circle, a set is formed with the users that have those interests. The set intersections that have at least one member with exactly those interests are shown in the diagram. The set intersections that do not have at least one member with those interests are not shown. To avoid any confusion as to context, our diagrams and questions made no reference to Twitter. All set names were changed, whilst keeping the real-world scenario: interests people share. The set names were chosen so that no two sets in any one Linear Diagram started with the same letter to reduce the potential for misreading and so that they were sufficiently short to avoid occlusion when presented as Hoop Diagrams with all our labelling variants. After prepilot experimentation by the project researchers, we settled on a suitable size of 6 sets, with 8 to 16 intersections, providing a variety of difficulties. After piloting this configuration in the first study, this was confirmed as an appropriate range of values to avoid ceiling and floor effects for correctness whilst permitting completion of the study in a reasonable time frame.

The study was performed using a web site developed on a Html/Javascript/Node.js platform. Prolific passed the prolific id, session id and study id to the initial web page. The initial page explained what the participants would do in the study, what information we would be recording, stated that they could stop the study at any time and asked for their consent to begin the study. If they gave their consent, those undertaking static conditions were presented with 4 training examples. Those undertaking the interactive conditions were presented with a short video (1 min 33 seconds) of the interactions followed by the 4 training examples. The training examples gave details of how to interpret the diagram, and, after a choice was made, some feedback on their answer was presented. See Figure 4 for an example of a training page with feedback after the choice was made. The training data sets were of varying sizes, starting small with 2 options, and ending with diagrams of the size typically seen in the study. Training Question 4 had 6 sets, 12 intersections and 4 options to choose from.

After the training phase, the study phase began. Questions were presented in a random order. For example, Figure 5 shows a question presented to a participant which is Question 9 in the list of questions. The order of the 4 options was randomized. Two attention check questions were presented, again randomly in the question sequence and with random answer order. These showed a diagram with the text "This is an attention check question." Three of the 4 options were standard looking answers similar to those in Figure 5 (although they did not relate to the diagram displayed, so could not be plausible alternative answers). The fourth option was the correct one which had the text "Choose this option". The second to last page was the demographics page. This asked questions about the participants Gender, Sight Difficulties, Age and an optional Any Comments free text box. The final page was a return link to Prolific with the completion code in the query string.

All diagrams, questions, anonymized data and code can be found at **https://www.eulerdiagrams.com/hoop/**. This web site also includes an interactive interface for creating Hoop Diagrams and Linear Diagrams.

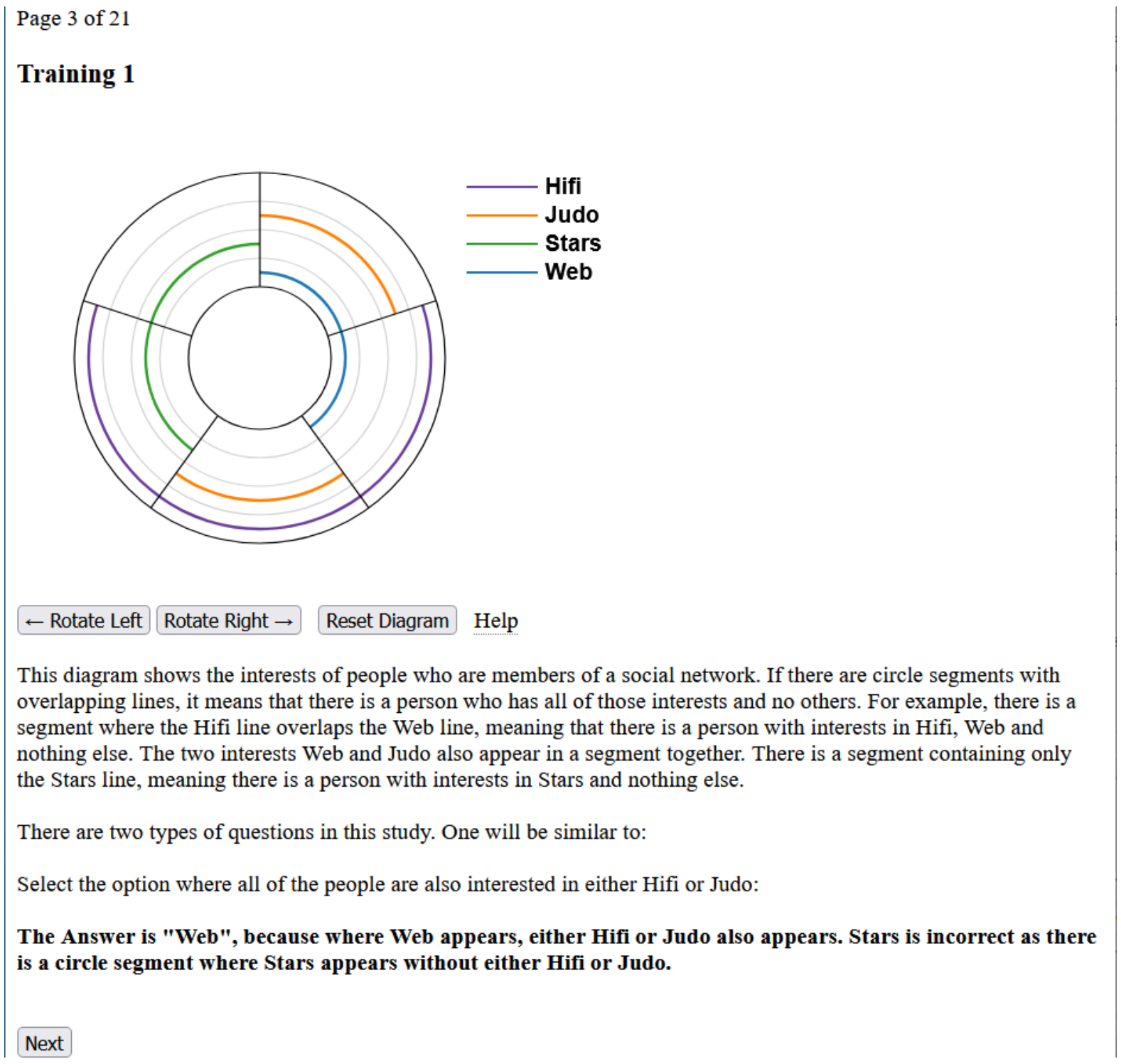

**Fig. 4.** Screenshot. First training example after the user makes a choice, so feedback on the correct answer is shown in bold. Previous to this, the participants were presented with the radio buttons “Stars” and “Web”. This is a dynamic Hoop Diagram.

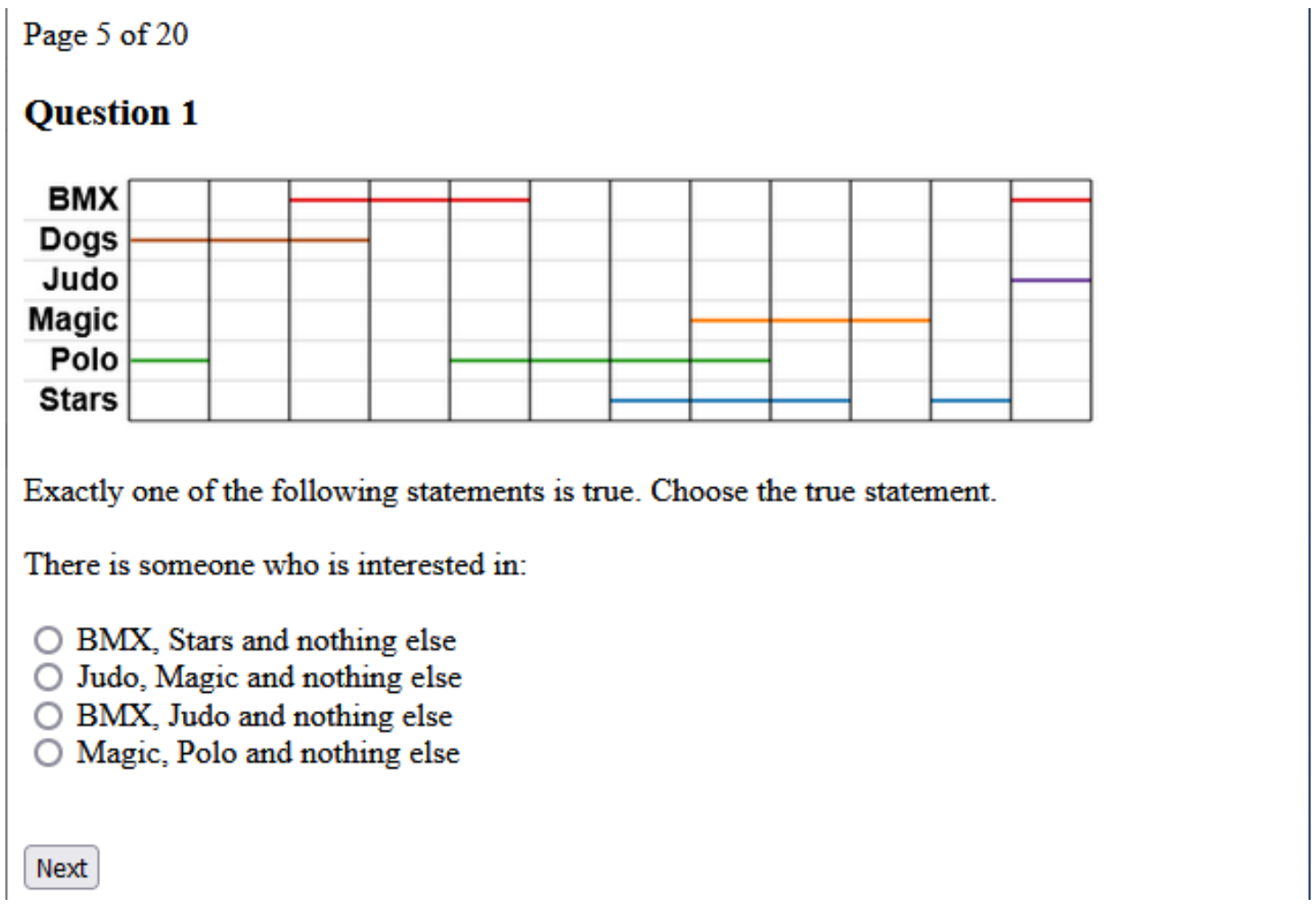

**Fig. 5.** A question as presented to the participants. This is a Static Linear Diagram. The correct answer is “BMX, Judo and nothing else”.

We stored anonymized participant and response data on a local server (University of Kent, UK).. For the interactive diagrams we stored hovers where changes in mouse position altered the highlights on the diagram, and clicks on the diagram and buttons.

Questions were of two types, with the same number of each in the study. The training had two of each type. The first question type was a set oriented question that was intended to require the close examination of entire sets to elicit the answer. It had the general form of: “Select the option where all of the people are also interested in X or Y” with single sets as answers. E.g., for Figure 6 the question was: “Select the option where all of the people are also interested in either Lego or Yoga:” with the options:

- Art
- Dogs
- Web
- Zumba

The correct answer was “Zumba”.

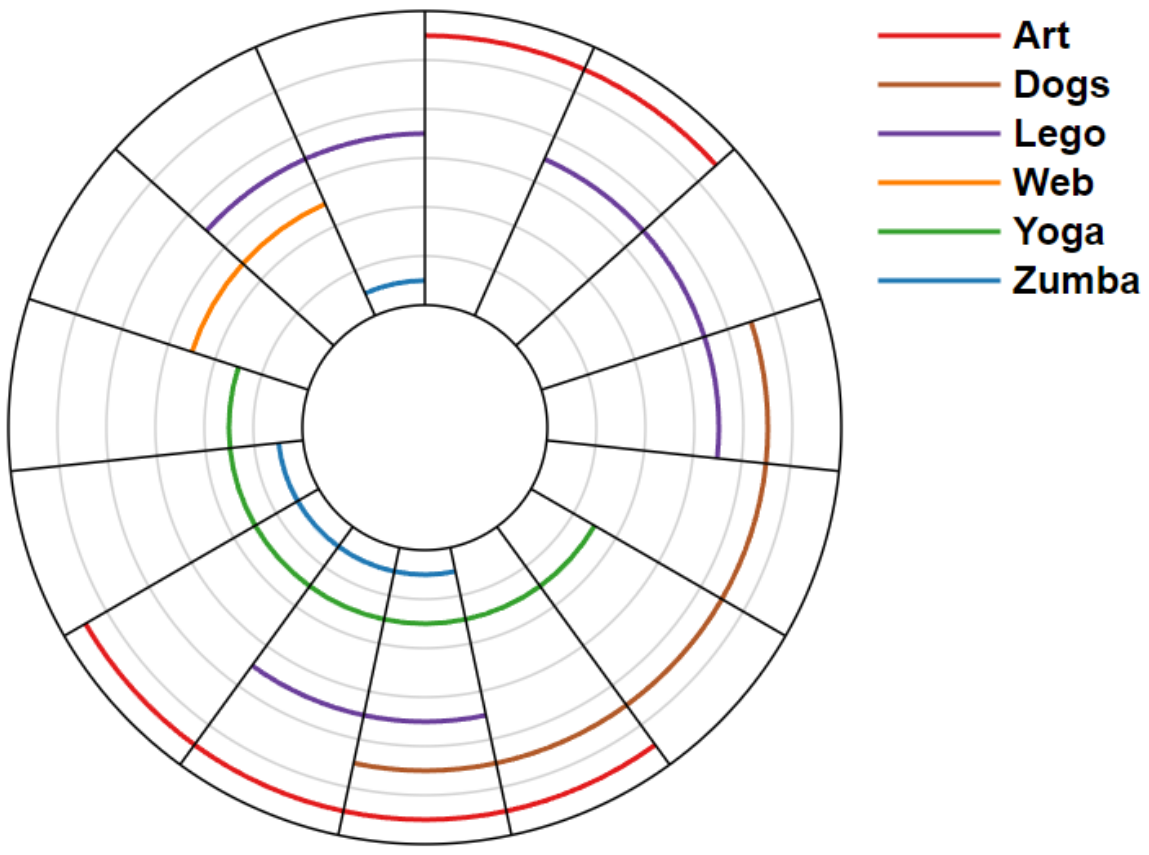

**Fig. 6.** Static Hoop Diagram for Question 4 from the final study.

The second question type was an intersection oriented question that intended to require the close examination of entire intersections to elicit the correct answer. It had the general form of: “Exactly one of the following statements is true. Choose the true statement. There is someone who is interested in:” with groups of sets as answers. For example, see Figure 5 which includes the question text and answer.

Our pilot had 20 participants. This was used to gauge task difficulty and to spot any technical issues with the studies, of which none were found and so no changes were made for the main study. The data from the pilot was discarded. The free text responses by participants was read immediately after all studies, pilots and main studies. No patterns or problems were identified from the free text entry.

We used Linear Diagrams drawn based on previous guidelines [20]. The design of Hoop Diagrams was consistent with the Linear Diagram design where possible. Labelling was via a legend. The interactive diagrams used the same diagram design as the static diagrams.

### 4.2 Statistical Methodology

We recorded two response variables: the time taken to answer a question (in seconds), and whether or not a participant correctly answered a question (a binary response). Taken across all participants and question instances, the mean correct responses provide a probability; namely, the probability that a randomly selected participant answers a randomly selected question correctly.

Multiple responses were collected from each participant (i.e. a time and success for each question, of which there were 12). These responses will therefore exhibit clustering, as the twelve responses from each individual are correlated with each other [17]. Use of standard statistical tests, such as ANOVA and chi-squared tests of goodness of fit, are insufficient, as such tests have an assumption of independence of observations. If we violate this assumption, then it can lead to overstated statistical significance, and underestimated standard errors. In other words, we may see a difference between groups where none exists.

To address this problem, a combination of generalized estimating equations (GEE) and generalized linear models (GLM) were used. The statistical software R was used for the analysis, primarily the package geepack [14]. The approach produces p-values, which can be interpreted in the usual way. In this instance, the p-value is the probability of obtaining the observed data given the coefficient of the independent variable in the model is 0. The models used had either a normal response variable (for time), or a binomial response variable (for accuracy). We considered the results to be significant if the p-value was less than 0.05.

## 5 Study results

The study compared four conditions: static Hoop Diagrams, static Linear Diagrams, dynamic Hoop Diagrams and dynamic Linear Diagrams in an attempt to answer H1, H2a and H2b. The goal was to provide guidance about the most effective diagram type for those visualizing set-based data.

**Table 1.** Hoop Diagram vs Linear Diagram. The green shaded cells indicate the value is significantly better than the other groups.

| Group | | Accuracy | | | Time (s) | | |
|---|---|---|---|---|---|---|---|
| | | **Overall** | **Set** | **Intersection** | **Overall** | **Set** | **Intersection** |
| **Hoop** | **Overall** | 0.798 | 0.6474 | 0.9025 | 51 | 55.4 | 48.5 |
| | **Interactive** | 0.779 | 0.6386 | 0.8825 | 53 | 57.8 | 51.1 |
| | **Static** | 0.808 | 0.6568 | 0.9236 | 48 | 52.9 | 45.7 |
| **Linear** | **Overall** | 0.757 | 0.6138 | 0.8662 | 46 | 49 | 44.1 |
| | **Interactive** | 0.762 | 0.6205 | 0.8656 | 48 | 52 | 46.2 |
| | **Static** | 0.752 | 0.6074 | 0.8668 | 43 | 46 | 42.1 |

To see which of Hoop Diagrams and Linear Diagrams are most suited to these sort of tasks, two analyses were performed: one with no reference to interaction (i.e. there were two groups, those who saw Hoop Diagrams, and those who saw Linear Diagrams), and a second analysis with four groups (i.e. interactive Hoop Diagrams, static Hoop Diagrams, interactive Linear Diagrams, and static Linear Diagrams). There was no split made on interactivity (i.e. interactive versus non-interactive) for reasons discussed in Section 6. There were 208 participants in the hoop group (107 interactive, 101 static), and 205 participants in the linear group (101 interactive, 104 static). The descriptive statistics are given in Table 1.

For the overall results, two generalized linear models were constructed with one independent variable (group) and either a binomial response variable (accuracy) or a normal response variable (time). For accuracy, the p-value is 0.061 (i.e. no significance), whereas for time there was a significant difference (p-value is 0.017), providing evidence that Linear Diagrams can be used more quickly than Hoop Diagrams. The effect size (Cohen’s d) is 0.099, which is small.

For the split by interaction as well as diagram type, two generalized linear models were constructed with one independent variable (group) and either a binomial response variable (accuracy) or a normal response variable (time). There were no significant differences between the groups for accuracy, whereas the linear static group were significantly faster than the other three (p-value is 0.0015). This improvement in time for linear static diagrams is evident in both question types (for intersection oriented questions, the p-value is 0.0074, whereas for set oriented questions, the p-value is 0.0046).

When looking at question type, for set oriented questions, there was no difference in accuracy between hoop and Linear Diagrams (p-value 0.253), but there is a difference in time: Linear Diagrams were faster (p-value 0.039). For intersection oriented questions, Hoop Diagrams improved accuracy (p-value 0.034) but were slower (p-value 0.046) than Linear Diagrams.

With regard to accuracy, our primary measure of diagram understandability, there is a significant result for Hoop Diagrams over Linear Diagrams in the case of intersection oriented questions. This leads us to the conclusion that when accurate analysis is the primary concern, and the questions are related to examination of intersections, rather than sets, Hoop Diagrams are preferred. For time, there are significant results favouring Linear Diagrams overall, as well as for static Linear Diagrams. This leads us to the conclusion that when time taken to complete an interpretation is the primary concern static Linear Diagrams are preferred.

With regards to our hypotheses, the differing statistical results for accuracy and time means we conclude that “H1: Hoop Diagrams are more understandable than Linear Diagrams” is false. For “H2a: Dynamic Linear Diagrams are more understandable than Static Linear Diagrams” has a timing result that indicates that this hypothesis is false. Finally, there is no evidence for “H2b: Dynamic Hoop Diagrams are more understandable than Static Hoop Diagrams” which we conclude to be false.

## 6 Exploration of Interaction Data

In this section, we explore what sort of interactions were used, and whether or not the used interactions were beneficial to participants (in terms of time and accuracy).

We make a distinction between necessarily intentional interactions (those which require a mouse click: reordering sets, reordering sectors, rotating diagrams, and resetting the diagram) and those which could be coincidental (highlighting). For the latter, for example, as a participant moves their cursor around the screen, they may highlight different sections of the diagram as an unintended consequence.

In what follows, we look only at the interactive group; that is, participants who had access to the interactive controls. Tables 2-4 show the proportion of question instances for which the indicated interactive element was used.

**Table 2.** Proportion using diagram highlighting

| | **Highlight (horizontal)** | | | **Highlight (vertical)** | | |
|---|---|---|---|---|---|---|
| | **Overall** | **Set** | **Intersection** | **Overall** | **Set** | **Intersection** |
| **Hoop** | 0.89 | 0.9 | 0.879 | 0.783 | 0.807 | 0.76 |
| **Linear** | 0.863 | 0.869 | 0.857 | 0.851 | 0.861 | 0.84 |

**Table 3.** Proportion using set reordering

| | **Reorder (horizontal)** | | | **Reorder (vertical)** | | |
|---|---|---|---|---|---|---|
| | **Overall** | **Set** | **Intersection** | **Overall** | **Set** | **Intersection** |
| **Hoop** | 0.27 | 0.292 | 0.247 | 0.264 | 0.294 | 0.234 |
| **Linear** | 0.225 | 0.259 | 0.191 | 0.294 | 0.342 | 0.246 |

**Table 4.** Proportion using rotation

| | **Rotate (left)** | | | **Rotate (right)** | | |
|---|---|---|---|---|---|---|
| | Overall | Set | **Intersection** | **Overall** | **Set** | **Intersection** |
| **Hoop** | 0.011 | 0.011 | 0.011 | 0.026 | 0.034 | 0.018 |
| **Linear** | 0.051 | 0.062 | 0.04 | 0.051 | 0.064 | 0.038 |

As discussed above, the intentional interactions are the two reorderings and the two rotations. Rotations appear to be used less frequently in Hoop Diagrams than in Linear Diagrams, and in general are used in at most 6.5% of diagram instances. Reordering is more common, with set reordering (ie. vertical reordering) used in 27.8% of question instances, and sector reordering (i.e. horizontal reordering) used in 24.8% of instances. In general though, these intentional interactions are not particularly common. For (possibly) unintentional highlighting, it is used in around 85% of question instances.

**Table 5.** Interaction accuracy. Green shaded cells show the highest accuracy.

| **Interactive Element** | **Accuracy whilst using the element** | **Accuracy whilst not using the element** |
|---|---|---|
| **Highlight - horizontal** | 0.7871 | 0.6993 |
| **Highlight - vertical** | 0.7865 | 0.6599 |
| **Reorder - horizontal** | 0.8395 | 0.7483 |
| **Reorder - vertical** | 0.8586 | 0.7371 |
| **Rotate - left** | 0.6438 | 0.7749 |
| **Rotate - right** | 0.6374 | 0.7762 |

We can also examine whether interactions appear to be helpful. This comparison is different to that given in Section 5. There, we were comparing the results of the groups against each other. Here, however, we are looking at average rates of success for participants who always had access to interactions, and either chose to use them or not. As the previous discussion on take-up shows (Tables 2-4), these rates vary widely amongst interactions, in some cases are very low or very high, and different participants used interactive elements on different questions. We therefore only give descriptive statistics, as the study presented was not designed to test these sorts of data.

As Table 5 shows, questions for which reordering was used had a higher absolute accuracy rate. As noted above, we cannot make any inferences about this apparent difference, but it is certainly suggestive that reordering is a useful interactive element. The same is true of highlighting, with the additional caveat that we cannot determine the participant's intention when using highlighting. By contrast, absolute accuracy rates are lower when using rotations, although rotations were used in only a limited number of cases (around 5% of question instances) so the evidence for any effect is weak.

## 7 Conclusions

We have introduced Hoop Diagrams, a new set visualization technique using an interactive diagram creating tool and empirical studies for diagram design and comparison against the closest existing set visualization technique, Linear Diagrams. The empirical results show that users answer questions on Linear Diagrams more quickly, but answer some questions more accurately with Hoop Diagrams. From a screen real estate perspective, Hoop Diagrams maintain a square aspect ratio, scaling in both horizontal and vertical directions, whereas Linear Diagrams only scale in the vertical direction with the number of sets and in the horizontal direction with the number of intersections. This horizontal scaling can quickly make Linear Diagrams too large for many displays. Hence, Hoop Diagrams may be preferred when presented on screen or paper.

Regards future possible work, the interaction functionality could be extended. The close relationship between Hoop Diagrams and Linear Diagrams could be exploited as Hoop Diagrams can be seen as Linear Diagrams with the ends of the diagram connected. Hence, it would be possible to show an animated transition between the two diagram types. A user could choose the diagram type depending on their personal preference. This would also allow the benefits of both diagram types for the user. So it would be possible to move from Linear to Hoop Diagram if the size of Linear Diagram means that it exceeds the space available for display, so allowing all the data to be shown at once for appropriate data sets.

Other future work includes developing interactivity aimed at improving users' performance at particular tasks. Being able to hide sets or intersections so that uninteresting areas of the diagram are no longer shown may simplify the display, so making analysis of data of interest easier. Similarly, using automatic highlighting to show potential areas of interest, such as sets or intersections might improve performance significantly.

**Acknowledgments.** Discussions about this work occurred at Dagstuhl Seminar 22462.